\pdfoutput=1
\RequirePackage{ifpdf}
\ifpdf 
\documentclass[pdftex]{sigma}
\else
\documentclass{sigma}
\fi

\begin{document}

\allowdisplaybreaks

\renewcommand{\thefootnote}{$\star$}

\renewcommand{\PaperNumber}{074}

\FirstPageHeading

\ShortArticleName{Ladder Operators for Lam\'e Spheroconal Harmonic Polynomials}

\ArticleName{Ladder Operators for Lam\'e Spheroconal \\
Harmonic Polynomials\footnote{This
paper is a contribution to the Special Issue ``Superintegrability, Exact Solvability, and Special Functions''. The full collection is available at \href{http://www.emis.de/journals/SIGMA/SESSF2012.html}{http://www.emis.de/journals/SIGMA/SESSF2012.html}}}

\Author{Ricardo M\'ENDEZ-FRAGOSO~$^{\dag \ddag}$ and Eugenio LEY-KOO~$^\ddag$}

\AuthorNameForHeading{R.~M\'endez-Fragoso and E.~Ley-Koo}

\Address{$^\dag$~Facultad de Ciencias, Universidad Nacional Aut\'onoma de M\'exico, M\'exico}
\EmailD{\href{mailto:rich@ciencias.unam.mx}{rich@ciencias.unam.mx}}
\URLaddressD{\url{http://sistemas.fciencias.unam.mx/rich/}}

\Address{$^\ddag$~Instituto de F\'{\i}sica, Universidad Nacional Aut\'onoma de M\'exico, M\'exico}
\EmailD{\href{mailto:eleykoo@fisica.unam.mx}{eleykoo@fisica.unam.mx}}

\ArticleDates{Received July 31, 2012, in f\/inal form October 09, 2012; Published online October 17, 2012}

\vspace{-1mm}

\Abstract{Three sets of ladder operators in spheroconal coordinates and their respective actions on Lam\'e spheroconal harmonic polynomials are presented in this article. The polynomials are common eigenfunctions of the square of the angular momentum operator and of the asymmetry distribution Hamiltonian for the rotations of asymmetric molecules, in the body-f\/ixed frame with principal axes.  The f\/irst set of operators for Lam\'e polynomials of a given species and a f\/ixed value of the square of the angular momentum raise and lower and lower and raise in complementary ways the quantum numbers $n_1$ and $n_2$ counting the respective nodal elliptical cones. The second set of operators consisting of the cartesian components $\hat L_x$, $\hat L_y$, $\hat L_z$ of the angular momentum connect pairs of the four species of polynomials of a chosen kind and angular momentum. The third set of operators, the cartesian components $\hat p_x$, $\hat p_y$, $\hat p_z$ of the linear momentum, connect pairs of the polynomials dif\/fering in one unit in their angular momentum and in their parities. Relationships among spheroconal harmonics at the levels of the three sets of operators are illustrated.}


\Classification{20C35; 22E70; 33C47; 33C80; 81R05}

\renewcommand{\thefootnote}{\arabic{footnote}}
\setcounter{footnote}{0}

\vspace{-3mm}

\section{Introduction}

One of our contributions to the Symposium on Superintegrability, Exact Solvability and Special Functions dealt with the topic of ``Symmetries and asymmetries in quantum systems conf\/ined by elliptical cones'', reviewing some of our results on the rotations of free asymmetric molecules \cite{LeyKooE/MendezFragosoR:20081,LeyKooE/MendezFragosoR:20082} and their extensions for other systems under conf\/inement \cite{MendezFragosoR/LeyKooE:2010,MendezFragosoR/LeyKooE:20112, MendezFragosoR/LeyKooE:2011}. While the free systems involve Lam\'e spheroconal harmonic polynomial eigenfunctions, the systems conf\/ined by elliptical cones involve quasi-periodic Lam\'e eigenfunctions represented as inf\/inite series, due to the breaking of the parity symmetry with respect the axis of the conf\/ining cone. Nevertheless, the methodology for the accurate and convergent evaluation of eigenvalues and eigenfunctions is the same.  The dif\/ferences reside in the boundary conditions, translating into recurrence relations of three terms and four terms for the expansion coef\/f\/icients, respectively; such relations can be cast into matrix forms which are of f\/inite size for the free systems \cite{LeyKooE/MendezFragosoR:20082} and of inf\/inite size for the conf\/ined systems \cite{MendezFragosoR/LeyKooE:2010,MendezFragosoR/LeyKooE:20112,MendezFragosoR/LeyKooE:2011} leading to the corresponding polynomial and inf\/inite series representations. The latter can be evaluated with suf\/f\/icient accuracy using f\/inite and large enough size matrices, testing for convergence along the way.

Concerning the conditions of exact solvability requiring polynomial solutions and ladder ope\-ra\-tors connecting them, the Lam\'e spheroconal harmonics for the free systems satisfy the f\/irst one and in this contribution three sets of ladder ope\-ra\-tors in the purely spheroconal formulation are reported.  It must be pointed out that one of these sets had been identif\/ied sometime ago \cite{LukachI:1973,LukachI/SmorodinskyY:1973,LukachI/SmorodinskyY:1970,PateraJ/WinternitzP:1973,PateraJ/WinternitzP:1976} and our recent work \cite{MendezFragosoR/LeyKooE:20112} in its section ``On Developing the Theory of Angular Momentum in Bases of Lam\'e Spheroconal Harmonics'' identif\/ied a second set.  However, it must also be admitted that in these treatments the connections with the spherical and cartesian harmonic bases were used.

The material in the manuscript is distributed in the following way: Section \ref{seceigenvalues} reviews brief\/ly the simultaneous separability of the square of the angular momentum operator and the asymmetry distribution Hamiltonian eigenvalue equations, the integration of the Lam\'e dif\/ferential equation in the respective elliptical cone coordinates, illustrating the lower eigenvalues and eigenfunctions for their matched products making up the Lam\'e spheroconal harmonic polynomials.  Section \ref{secrln1n2} identif\/ies the raising and lowering actions on the numbers $n_1$ and $n_2$ counting the nodal elliptical cones in the neighboring spheroconal harmonic polynomials as their respective eigenvalues $h^A _{n_1} (k^2 _1)$ and $h^B _{n_2} (k^2 _2)$ are complementarily increased and decreased, which follows as a corollary to the tridiagonal matrix diagonalization method of solution, for chosen values of~$\ell$ and species~$AB$ of the polynomials. Section~\ref{secamtheory} identif\/ies the shifting actions of the~$\hat L _x$,~$\hat L _y$ and~$\hat L _z$ operators on the individual Lam\'e polynomials of species $A$ and $B$, respectively, and on their matched products of species $AB$ for each f\/ixed value of~$\ell$. Section \ref{sectionshift} identif\/ies the corresponding actions of the $\hat p_x$, $\hat p_y$ and $\hat p_z$ operators on an initial harmonic polynomial~$\ell AB$, raising and lowering the angular momentum by one unit $\ell ' = \ell \pm 1$ into f\/inal polynomials $\ell ' A' B'$ of opposite parity and dif\/ferent species.  Section~\ref{secdiscu} discusses the main results of this work, including relationships among the spheroconal harmonics in the three levels of the identif\/ied operators, as well as connections with other works in the literature.

\section[Eigenvalues and eigenfunctions of $\hat L^2$ and $\hat H ^*$]{Eigenvalues and eigenfunctions of $\boldsymbol{\hat L^2}$ and $\boldsymbol{\hat H ^*}$}\label{seceigenvalues}

Kramers and Ittmann \cite{KramersHA/IttmannGP:1929} pioneered the quantum mechanical study of the rotations of asymmetric molecules using spheroconal coordinates, showing the separability of the eigenvalue equations for the square of the angular momentum and the complete Hamiltonian
\begin{gather}
\hat H = \frac{1}{2} \left( \frac{\hat L^2 _x}{I _1} + \frac{\hat L^2 _y}{I _2} + \frac{\hat L^2 _z}{I _3} \right)
\label{completeham}
\end{gather}
in the body-f\/ixed frame with principal axes, involving the moments of inertia $I_1 \leq I_2 \leq I_3$. The separated equations are of the Lam\'e type, and its polynomial solutions are exact in principle, but could not be numerically implemented for higher excited states at that time.  Consequently, the study of the rotations of asymmetric molecules developed along the perturbation theory route taking the symmetric prolate and oblate molecule spherical harmonic solutions as the non-perturbed starting point \cite{KronigRdeL/RabiJJ:1927,KrotoHW:1975,LutgemeierF:1926,ReicheF/RademacherH:1926,WitmerEE:1926}.

Patera and Winternitz used an alternative energy operator
\begin{gather*}
E = -4 \big( \hat L^2 _1 + r \hat L^2 _2 \big), \qquad 0<r<1, 
\end{gather*}
in their work 
\cite{PateraJ/WinternitzP:1973}.  On the other hand, Pi\~na \cite{PinaE:1999} and Vald\'ez and Pi\~na \cite{ValdezMT/PinaE:2006} used the alternative parametrization of the original Hamiltonian of equation~(\ref{completeham}), in the form
\begin{gather*}
\hat H = \frac{1}{2} Q \hat L ^2 + \frac{1}{2}P \big( e_1 \hat L ^2 _x +  e_2 \hat L ^2 _y + e_3 \hat L ^2 _z \big), 
\end{gather*}
where
\begin{gather*}
Q = \frac{1}{3} \left( \frac{1}{I_1} + \frac{1}{I_2} + \frac{1}{I_3}  \right)
\end{gather*}
is the average of the three inverses of the moments of inertia characterizing a spherical top,
\begin{gather*}
P^2 = \frac{2}{9} \left[ \left( \frac{1}{I_1} - \frac{1}{I_2} \right)^2 + \left( \frac{1}{I_1} - \frac{1}{I_3} \right)^2 + \left( \frac{1}{I_2} - \frac{1}{I_3} \right)^2 \right] 
\end{gather*}
measures the magnitude of the asymmetry of the molecule and
\begin{gather}
\hat H^* = \frac{1}{2} \big( e_1 L ^2 _x +  e_2 L ^2 _y + e_3 L ^2 _z \big) \label{hstar}
\end{gather}
is the asymmetry distribution Hamiltonian, with parameters $e_1$, $e_1$ and $e_3$ such that
\begin{gather} \label{eiparameters}
 e_1 \geq  e_2  \geq e _3,\qquad
 e_1 + e_2  + e _3 = 0,\qquad
 e ^2 _1 +  e ^2 _2 + e^2 _3  = \frac{3}{2}.
\end{gather}

The inequalities follow from those for the moments of inertia, the vanishing of their sum ref\/lects the zero trace of the matrix of the inverses of the moments of inertia after the term in~$Q$ is separated, and the coef\/f\/icient in~$P^2$  and the sums of the squares are correlated.  Only one of the parameters $e_i$ is independent, the set of~$I_1$, $I_2$, $I_3$ being replaced by $Q$, $P$, $e_i$.

\looseness=-2
Both of our works \cite{LeyKooE/MendezFragosoR:20081, LeyKooE/MendezFragosoR:20082} used the parametrizations of equations~(\ref{hstar}) and~(\ref{eiparameters}) with a single independent parameter $e_i$, in analogy with that in equation~(\ref{completeham}). At that time we were not aware of~\cite{PateraJ/WinternitzP:1973}. Not surprisingly, their equivalence can be established and our respective methodologies also overlap.  In fact, our solution using the spherical canonical bases to construct the matrix representation of~$\hat H^*$ and determine its eigenvalues and eigenfunctions by diagonalization~\cite{LeyKooE/MendezFragosoR:20082} is related to Sections~2~and~3 in~\cite{PateraJ/WinternitzP:1973}.  On the other hand, our construction of the Lam\'e eigenvalues and polynomials also by diagonalization of the matrices representing the recurrence relations for the series expansions in the Jacobi elliptic function representation of the spheroconal coordinates~\cite{LeyKooE/MendezFragosoR:20082}, has as well points of contact with parts of Section~6 in~\cite{PateraJ/WinternitzP:1973}. We have just become aware of some additional references~\cite{LukachI:1973,LukachI/SmorodinskyY:1970} and~\cite{LukachI/SmorodinskyY:1973} of that period, whose titles describe the overlap with this contribution. Specif\/ically, the role of the~$D_2$ group of rotations by~$\pi$ around the cartesian coordinate axes as a f\/inite symmetry group is explicitly recognized in~\cite{PateraJ/WinternitzP:1976} as the missing ingredient for the complete characterization of the dif\/ferent species of Lam\'e spheroconal harmonics.

We refer the reader to \cite{LeyKooE/MendezFragosoR:20082,PinaE:1999,ValdezMT/PinaE:2006} for the details or equations in what follows.  The transformation equations between spheroconal and cartesian coordinates are
\begin{gather}
 x = r \operatorname{dn} \big(\chi _1 |k ^2 _1 \big)\operatorname{sn} \big(\chi _2 |k ^2 _2 \big), \qquad
 y = r \operatorname{cn} \big(\chi _1 |k ^2 _1 \big)\operatorname{cn} \big(\chi _2 |k ^2 _2 \big), \nonumber\\
 z = r \operatorname{sn} \big(\chi _1 |k ^2 _1 \big)\operatorname{dn} \big(\chi _2 |k ^2 _2 \big)  \label{transcoord}
\end{gather}
in terms of Jacobi elliptic functions $\operatorname{sn} \big(\chi | k^2 _i\big)$, $\operatorname{cn}  \big(\chi | k^2 _i\big)$ and $\operatorname{dn}  \big(\chi | k^2 _i\big)$ with matching parameters $k^2 _1 + k^2 _2 = 1$  \cite{AbramowitzM/StegunIA:1965,MorsePM/FeshbachH:1953}. The elliptic functions are related as
\begin{gather}
\operatorname{cn}^2 \big(\chi  | k^2 \big)   =   1 - \operatorname{sn}^2 \big(\chi  | k^2 \big),\qquad
\operatorname{dn}^2 \big(\chi  | k^2 \big)   =   1 - k^2 \operatorname{sn}^2 \big(\chi  | k^2 \big),  \label{defcndn}
\end{gather}
and their derivatives are
\begin{gather}
\frac{d}{d \chi} \operatorname{sn} \big(\chi  | k^2 \big)   =    \operatorname{cn} \big(\chi  | k^2 \big) \operatorname{dn} \big(\chi  | k^2 \big),  \qquad
\frac{d}{d \chi} \operatorname{cn} \big(\chi  | k^2 \big)   =   - \operatorname{sn} \big(\chi  | k^2 \big) \operatorname{dn} \big(\chi  | k^2 \big), \nonumber \\
\frac{d}{d \chi} \operatorname{dn} \big(\chi  | k^2 \big)   =   - k^2 \operatorname{sn} \big(\chi  | k^2 \big) \operatorname{cn} \big(\chi  | k^2 \big).\label{derivesellip}
\end{gather}
The transformation equations~(\ref{transcoord}) and equations~(\ref{defcndn}) allow the identif\/ication of~$r$ as the radial spherical coordinate, $\chi _1$ as an elliptical cone coordinate with axis along the $x$-axis and~$\chi_2$ as an elliptical cone coordinate with axis along $z$-axis. Their respective domains are $-\pi < \operatorname{am} (\chi _i) < \pi$ and $-\pi/2 < \operatorname{am} (\chi _j) < \pi/2$ for $k^2 _i > k^2 _j$. Since $k^2 _1$ is always associated with variable $\chi_1$ and $k^2 _2$ with $\chi _2$, from now on in this manuscript, we simplify the notation as $\big(\chi _i | k^2 _i\big) \to (\chi _i)$, for the sake of space saving.

The transformation equations (\ref{transcoord}) and the equations (\ref{derivesellip}) lead in turn to the scale factors
\begin{gather}
h_r = 1  , \qquad  h_{\chi} = h_{\chi _1} = h_{\chi _2} = r \sqrt{1 - k^2 _1 \operatorname{sn} ^2 (\chi _1 ) - k^2 _2 \operatorname{sn} ^2 (\chi _2 ) } \label{scalefac}
\end{gather}
and to the Laplace operator
\begin{gather*}
\nabla ^2  =  \frac{1}{r^2} \frac{\partial}{\partial r} \left( r^2 \frac{\partial}{\partial r} \right) + \frac{1}{{h_\chi ^2 }}\left( {\frac{{\partial ^2 }}{{\partial \chi _1^2 }} + \frac{{\partial ^2 }}{{\partial \chi _2^2 }}} \right)  .
\end{gather*}
From the latter the identif\/ication of the square of the angular momentum is immediate:
\begin{gather}
\hat L^2  =  - \frac{\hbar ^2 r^2}{{h_\chi ^2 }}\left( {\frac{{\partial ^2 }}{{\partial \chi _1^2 }} + \frac{{\partial ^2 }}{{\partial \chi _2^2 }}} \right)  . \label{lsquare}
\end{gather}

The cartesian components of the angular momentum vector itself can be found in the Appen\-dix of~\cite{LeyKooE/MendezFragosoR:20082}, and allow the construction of the explicit form of the asymmetry distribution Hamiltonian
\begin{gather}
\hat H^*  = - \frac{\hbar ^2 r^2}{{2h_\chi ^2 }}\left( {\left( {e_1  - (e_1  - e_2 )\operatorname{sn}^2 (\chi _2 )} \right)\frac{{\partial ^2 }}{{\partial \chi _1^2 }} + \left( {e_3  - (e_2  - e_3 )\operatorname{sn}^2 (\chi _2 )} \right)\frac{{\partial ^2 }}{{\partial \chi _2^2 }}} \right). \label{hstarspheroconal}
\end{gather}
Since $\hat L ^2$ commutes with $\hat H^*$, as easily tested using equation~(\ref{hstar}), both operators admit common eigenfunctions with the factorizable form
\begin{gather}
\Psi (\chi _1 , \chi _2) = \Lambda (\chi _1) \Lambda (\chi _2) \label{factorizablefunc}
\end{gather}
for which the eigenvalue equations of the operators of equation~(\ref{lsquare}) and (\ref{hstarspheroconal}), with eigenvalues $\ell (\ell + 1)$ and $E^*$ become separable. Indeed by using the angular scale factors of equation~(\ref{scalefac}), the separability can be accomplished provided the geometric parameters in equations~(\ref{transcoord}) and the dynamic parameters in equation~(\ref{hstarspheroconal}) are connected by
\begin{gather*}
k ^2 _1 = \frac{e_2 - e_3}{e_1 - e_3}  , \qquad k ^2 _2 = \frac{e_1 - e_2}{e_1 - e_3}.
\end{gather*}
This means that the specif\/ic spheroconal coordinate system is tailored to the asymmetry distribution of the molecule.

The next result is that each of the factors in equation~(\ref{factorizablefunc}) satisf\/ies the Lam\'e equation \cite{NivenWD:1891,VolkmerH:2010,WhittakerET/WatsonGN:1927}
\begin{gather}
\left[ - \frac{{d^2  }}{{d\chi _i^2 }} +  {\ell (\ell  + 1)k_i^2 \operatorname{sn} ^2 (\chi _i ) } \right] \Lambda _i (\chi _i)  = h_i  \Lambda _i (\chi _i) \label{lameeq}
\end{gather}
for  $i =1,2$, with the respective eigenvalues
\begin{gather}
 h_1  = - \frac{{\ell (\ell  + 1)e_3 }}{{e_1  - e_3 }} + \frac{{2E^* }}{{e_1  - e_3 }}
 , \qquad
 h_2  =   \frac{{\ell (\ell  + 1)e_1 }}{{e_1  - e_3 }} - \frac{{2E^* }}{{e_1  - e_3 }}  .
 \label{hcombinations}
\end{gather}
Their combinations{\samepage
\begin{gather}
 h_1  + h_2  =  \ell (\ell  + 1)  , \qquad  e_1 h_1  + e_3 h_2  =   2E^*   , \label{eigenvalrelation}
 \end{gather}
provide the eigenvalues of $\hat L^2$ and $\hat H^*$, respectively.}

Next, the solutions of the Lam\'e equation~(\ref{lameeq}) are also reviewed.  The derivative equations~(\ref{derivesellip}) for the elliptic functions, represented by their initial letters allow the identif\/ication of $\emph{1}$, $s$, $c$, $d$, $dc$, $ds$, $sc$, $dcs$ as removable singularity factors in  the dif\/ferential equation.  The dependence of both operators $\hat L^2$ and $\hat H^*$ on the squares of $\hat L _x$, $\hat L_y$ and $\hat L _z$ guarantee that their common eigenfunctions, equation~(\ref{factorizablefunc}), have def\/inite inversion parities under $x \to -x$, $y \to -y$, $z \to -z$.  Correspondingly, the Lam\'e functions get classif\/ied into two kinds, each with four species sharing the same number of even or odd factors in the removable singularity factors.  The Lam\'e functions in the respective elliptic cone coordinates $\chi _1$ and $\chi _2$  in equation~(\ref{factorizablefunc}) must be matched according to their  singularity factors  $A (\chi _1)$ and $B (\chi _2)$, and the respective parities from equations~(\ref{transcoord}), as illustrated in Table~\ref{clasifspheroconalhar}.

\begin{table}[t]\centering
\caption{Matching Lam\'e polynomials according to species and parities.}
\label{clasifspheroconalhar}
\vspace{1mm}

\begin{tabular}{  c | c c c c c c c c }
A &  {\em 1}  &  $d$  & $c$ & $s$ & $dc$ &  $ds$  &  $cs$  & $dcs$ \\
B &  {\em 1}  &  $s$  & $c$ & $d$ & $sc$ &  $sd$ &  $cd$  & $scd$ \\ \hline
AB & {\em 1} & $x$ & $y$ & $z$ & $xy$ & $xz$ & $yz$ & $xyz$  \\ \hline
$\Pi _x$ & + & $-$ & + & + & $-$ & $-$ & + & $-$ \\
$\Pi _y$ & + & + & $-$ & + & $-$ & + & $-$ & $-$ \\
$\Pi _z$ & + & + & + & $-$ & + & $-$ & $-$ & $-$
\end{tabular}
\end{table}

There are two kinds of Lam\'e spheroconal harmonic functions distinguished by their overall parity $(-)^{\ell} = \Pi _x \Pi _y \Pi _z$.  For each value of $\ell$ each kind may contain four dif\/ferent species characterized by the respective singularity factors.

The solutions of the Lam\'e dif\/ferential equation~(\ref{lameeq}) have the general form
\begin{gather}
\Lambda ^A (\chi _i ) = A(\chi _i )\sum\limits_{s = 0}^{N_{\max }^A } {a_s^A  \operatorname{sn}^{2s} (\chi _i )} . \label{lamepoly}
\end{gather}
as a series of even-power of the elliptical $\operatorname{sn} (\chi _i)$ function, where the expansion coef\/f\/icients satisfy three-term recurrence relations, which the interested reader may see in equations~(32)--(39) of~\cite{LeyKooE/MendezFragosoR:20082}, for the eight dif\/ferent species~$A$.  For the lower~$\ell$ eigenstates the eigenvalues $h^A \big(k^2 _i\big)$ and ratios of the expansion coef\/f\/icients~$a^A _s / a^A _0$ can be evaluated in a straightforward way from equation~(\ref{lameeq}), and the linear independence of the powers of $\operatorname{sn} (\chi _i )$. The corresponding results are illustrated in Table~\ref{eigenfunc2} for the individual Lam\'e polynomials and in Table~\ref{eigenenergy2} for the matched spheroconal harmonics.

For higher values of $\ell$ it is more practical to cast the recurrence relations into eigenvalue problem matrix forms and to obtain from their diagonalizations the respective eigenvalues and expansion coef\/f\/icients.  The sizes of the matrices are as follows for $\ell$ even:
\begin{gather*}
N^{\textit 1} _{\max} = \frac{\ell}{2} +1 \qquad \textrm{and} \qquad N^{sc} _{\max} = N^{sd} _{\max} = N^{cd} _{\max} = \frac{\ell}{2}  ,
\end{gather*}
and for $\ell$ odd:
\begin{gather*}
N^{s} _{\max} = N^{c} _{\max} = N^{d} _{\max} = \frac{\ell+1}{2} \hspace{1cm} \textrm{and} \hspace{1cm} N^{scd} _{\max} = \frac{\ell -1}{2} .
\end{gather*}
In both cases the total dimensions are $2\ell +1$, corresponding to the number of independent spheroconal harmonic polynomials for each value of $\ell$.  For each species there are $N^A _{\max}$ dif\/ferent eigenvalues which can be ordered according to their increasing values
\begin{gather}
 h^A _1 < h^A _2 < h^A _3 < \cdots < h^A _i < \cdots < h^A _{ N^A _{\max} -1} < h^A _{ N^A _{\max}} , \label{haorder}
\end{gather}
or their decreasing values
\begin{gather}
 h^B _{N_{\max}} > h^B _{N_{\max} -1}> \cdots > h^B _j > \cdots > h^B _3 > h^B _2 > h^B _1 . \label{hborder}
\end{gather}
Obviously, the expansion coef\/f\/icients in equation~(\ref{lamepoly}) depend on the specif\/ic eigenvalues $a_s (h^A _i)$, $a_s (h^B _j)$. The f\/irst of equations~(\ref{eigenvalrelation}) suggests the matching of the smaller eigenvalues of one species with the larger ones of the companion species for the construction of the complete set of spheroconal harmonics $\ell AB ij$.

Let us also analyze the number of nodes in the respective factors of Lam\'e functions in equa\-tion~(\ref{lamepoly}), $n^A$ and $n^p$. For the singularity removing factors their specif\/ic values are as follows:
\begin{gather}
 n^{\emph{1}} = n^d =0  , \qquad n^c = 0 \ {\rm or} \  1  , \qquad n^s = 1 , \nonumber \\
 n^{dc} = 0 \  {\rm or} \  1 , \qquad n^{ds} = 1 , \qquad n^{cs }= 1 \  {\rm or } \  2, \qquad n^{dcs} = 1 \  {\rm or} \  2  . \label{numbernodes}
\end{gather}
The polynomial of degree $2 N ^A _{\max}$ in $\operatorname{sn} (\chi)$ has at the most $N^A _{\max}$ zeros $\operatorname{sn} ^2 (\chi _p)$, appearing by pairs with opposite signs $\pm \operatorname{sn} (\chi _p)$ with the possible numbers
\begin{gather}
n^p =  0,2,4 , \dots, 2(N_{\max} -1) \label{nodesnp}
\end{gather}
counting the symmetric sheets of the nodal elliptical cones.

Therefore, the total number of nodes for $\Lambda ^{\ell A} _{n_1} (\chi _1)$ Lam\'e polynomial is
\begin{gather}
n_1 = n^A + n^p = 0,1,2,\dots,\ell \label{nodesn1}
\end{gather}
and likewise for its matching companion $\Lambda ^{\ell B} _{n_2} (\chi _2)$
\begin{gather}
n_2 = n^B + n^p = 0,1,2,\dots ,\ell. \label{nodesn2}
\end{gather}

Correspondingly, the spheroconal harmonic functions from the product of the pairs of matched Lam\'e polynomials exhibit the possible numbers
\begin{gather}
 n^A + n^B = 0,1,2,3 \label{naplusnb}
\end{gather}
counting the shared nodal cartesian planes, as limiting cases of nodal elliptical cones, for the respective species
\begin{gather*}
AB = \emph{1} ; x,y,z ; xy ,xz, yz ; xyz  .
\end{gather*}

In turn,
\begin{gather}
n _1 + n _2 = \ell \label{totalnodes}
\end{gather}
counts the total number of nodes characterizing any harmonic function of order $\ell$.

The ordering indices $i$, $j$ for the eigenvalues in equations (\ref{haorder}) and (\ref{hborder}) can be replaced with the quantum numbers $n_1$ and $n_2$. This is completely justif\/ied by recognizing that the eigenvalues with matching quantum numbers, according to equation~(\ref{totalnodes}), are also matched to satisfy
\begin{gather}
h^{\ell A} _{n _1} + h^{\ell B} _{n _2} = \ell (\ell + 1) \label{eigenvaltotal}
\end{gather}
as required by equation~(\ref{eigenvalrelation}).

Table~\ref{eigenfunc2} illustrates the eigenvalues $h^{\ell A} _{n_i}$ of the Lam\'e functions for the successive values of $\ell = 0,1,2,\dots $, and the respective species with the singularity removing factors. The eigenvalues and their quantum numbers $n_1 = 0,1,2,\dots $ are listed in increasing order. The associated polynomial factors in the eigenfunctions, equation~(\ref{lamepoly}), are also included.

\begin{table}[t!]\centering
\caption{Lam\'e eigenvalues $h^{\ell A} _{n_1}$ and polynomials.}
\label{eigenfunc2}

\vspace{1mm}

\begin{tabular}{c | l | l | l}
 & $A$ & $h^A _{n_i}$\bsep{2pt} & Polynomial \\ \hline \hline
$\ell =0$ & 1 & $h^{\emph{1}} _0 =0$\tsep{2pt} & 1 \\ \hline
$\ell =1$ & $d$ & $h^d _0 = k^2 _i$\tsep{2pt} & 1 \\
          & $c$ & $h^c _1 = 1$      &  1 \\
          & $s$ & $h^s _1 = 1+k^2 _i$ & $1$ \\ \hline
 $\ell =2$ & $dc$ & $h^{dc} _1 = 1 + k^2 _i$\tsep{2pt}  & 1 \\
          & $ds$ & $h^{ds} _1 = 1 + 4k^2 _i$ & 1 \\
          & $cs$ & $h^{cs} _2 = 4 + k^2 _i$  & 1 \\
          & $\emph{1}$ & $h^{\emph{1}} _0 = 2(1 + k^2 _i) - 2\sqrt{1-k^2 _i ( 1- k^2 _i)}$ & $ 1-  \frac{h^{\emph{1}} _0}{2} \operatorname{sn}^2 (\chi _i) $ \\
          & $\emph{1}$ & $h^{\emph{1}} _2 = 2(1 + k^2 _i) + 2\sqrt{1-k^2 _i ( 1-  k^2 _i)}$ & $1-  \frac{h^{\emph{1}} _2}{2} \operatorname{sn}^2 (\chi _i) $ \\ \hline
 $\ell =3$ & $dcs$ & $h^{dcs} _2 = 4(1 + k^2 _i)$\tsep{2pt} & 1 \\
          & $d$ & $h^{d} _0 = (5k^2 _i +2) - 2\sqrt{4k^4 _i - k^2 _i +1}$ & $1+  \frac{k^2 _i - h^{d} _0 }{2} \operatorname{sn}^2 (\chi _i) $ \\
          & $d$ & $h^{d} _2 = (5k^2 _i +2) + 2\sqrt{4k^4 _i - k^2 _i +1}$ & $1+  \frac{k^2 _i - h^{d} _2 }{2} \operatorname{sn}^2 (\chi _i) $ \\
          & $c$ & $h^{c} _1 = (5 + 2k^2 _i) - 2\sqrt{4 - k^2 _i (1 - k^2 _i)}$ &  $1+  \frac{1 - h^{c} _1 }{2} \operatorname{sn}^2 (\chi _i) $ \\
          & $c$ & $h^{c} _3 = (5 + 2k^2 _i) + 2\sqrt{4 - k^2 _i (1 - k^2 _i)}$ &  $1+  \frac{1 - h^{c} _3 }{2} \operatorname{sn}^2 (\chi _i) $ \\
          & $s$ & $h^{s} _1 = 5(1 + k^2 _i) - 2\sqrt{4k^4 _i - 7k^2 _i +4}$ & $ 1+  \frac{1 +k^2 _i - h^{s} _1 }{6} \operatorname{sn}^2 (\chi _i) $ \\
          & $s$ & $h^{s} _3 = 5(1 + k^2 _i) + 2\sqrt{4k^4 _i - 7k^2 _i +4}$ &  $1+  \frac{1 +k^2 _i - h^{s} _3 }{6} \operatorname{sn}^2 (\chi _i) $ \\ \hline
$\ell =4$ & $\emph{1}$ & $h^3 - 20 (1+ k^2 _i)h^2   $\tsep{2pt} & $ 1 + \frac{a_1}{a_0}  \operatorname{sn}^2 (\chi _i) + \frac{a_2}{a_0} \operatorname{sn}^4 (\chi _i) $\\
                & & \hspace{1cm} $ {}+[64(1+ k^2 _i) + 208 k^2 _i ]h  $ & $a_1 = - \frac{h^{\emph{1}} _{n_i}}{2} a_0 $; $a_2 = \frac{7k^2 _i h^{\emph{1}} _{n_i}}{16(1+k^2 _i) - h^{\emph{1}} _{n_i}} a_0 $ \\
          & & \hspace{2cm} $  {}-640 k^2 _i(1 + k^2 _i) =0$ & The roots of the polynomial \\
          & & & are $h^{\emph{1}} _0$, $h^{\emph{1}} _2$, $h^{\emph{1}} _4$ \\
          & $dc$ & $h^{dc} _1 = 5(1 + k^2 _i) - 2\sqrt{4 + k^2 _i + 4k^4 _i}$  & $ 1- \frac{h^{dc} _1 - (1 + k^2 _i)}{2} \operatorname{sn}^2 (\chi _i) $ \\
          & $dc$ &$h^{dc} _3 = 5(1 + k^2 _i) + 2\sqrt{4 + k^2 _i + 4k^4 _i}$  & $ 1- \frac{h^{dc} _3 - (1 + k^2 _i)}{2} \operatorname{sn}^2 (\chi _i) $ \\
          & $ds$ & $h^{ds} _1 = 5(1 + 2k^2 _i) - 2\sqrt{4 - 9k^2 _i (1 - k^2 _i})$  & $ 1- \frac{h^{ds} _1 - (1 + 4k^2 _i)}{6} \operatorname{sn}^2 (\chi _i) $ \\
          & $ds$ & $h^{ds} _3 = 5(1 + 2k^2 _i) + 2\sqrt{4 - 9k^2 _i (1-k^2 _i)}$  & $ 1- \frac{h^{ds} _3 - (1 + 4k^2 _i)}{6} \operatorname{sn}^2 (\chi _i) $ \\
          & $cs$ & $h^{cs} _2 = 5(2 + k^2 _i) - 2\sqrt{9 - 9k^2 _i + 4k^4 _i}$  & $ 1- \frac{h^{cs} _2 - (4 + k^2 _i)}{6} \operatorname{sn}^2 (\chi _i) $ \\
          & $cs$ & $h^{cs} _4 = 5(2 + k^2 _i) + 2\sqrt{9 - 9k^2 _i + 4k^4 _i}$  & $ 1- \frac{h^{cs} _4 - (4 + k^2 _i)}{6} \operatorname{sn}^2 (\chi _i) $ \\
\end{tabular}
\end{table}

Table \ref{eigenenergy2} illustrates the matching of Lam\'e eigenvalues and polynomials to obtain the spheroconal harmonics $\ell A B n_1 n_2$ and their eigenenergies $2E^*$ from equation~(\ref{eigenvalrelation}), written in terms of the asymmetry distribution parameter $e_i$. The readers may check the ordering of the individual Lam\'e eigenvalues, correlated with the increasing of the number of nodes in the polynomials in Table~\ref{eigenfunc2}. The matching according to equations~(\ref{totalnodes}) and~(\ref{eigenvaltotal}), and the ordering of increasing energies can be appreciated in Table~\ref{eigenenergy2}.

\begin{table}[t]\centering
\caption{Eigenenergies $2E^*$ in units of $\hbar ^2$ for Lam\'e spheroconal harmonics $\ell AB n_1 n_2$.}
\label{eigenenergy2}

\vspace{1mm}
\begin{tabular}{c | l l | l l | l}
 & $A$ & $B$ & $n_1$ & $n_2$ & $2E^*$ \\ \hline \hline
$\ell =0$ & $\emph{1}$ & $\emph{1}$ & 0 & 0 & 0 \\ \hline
$\ell =1$ & $d$ & $s$ & 0 & 1 & $-e_1$ \\
                & $c$ & $c$ & 1 & 0 & $-e_2$ \\
                & $s$ & $d$ & 1 & 0 & $-e_3$ \\ \hline
$\ell =2$ & $\emph{1}$ & $\emph{1}$ & 0 & 2 & $-3$ \\
                & $dc$ & $sc$ & 1 & 1 & $3e_3$ \\
                & $ds$ & $sd$ & 1 & 1 & $3e_2$ \\
                & $cs$ & $cd$ & 2 & 0 & $3e_1$ \\
                & $\emph{1}$ & $\emph{1}$ & 2 & 0 & 3 \\ \hline
 $\ell =3$ & $d$ & $s$ & 0 & 3 & $-3e_1- 3\sqrt{5 - 4e ^2 _1} $\tsep{3pt} \\
                & $c$ & $c$ & 1 & 2 & $-3 e_2 - 3\sqrt{5 - 4e ^2 _2}$ \\
                & $s$ & $d$ & 1 & 2 & $-3 e_3 - 3\sqrt{5 - 4 e ^2 _3}$ \\
                & $dcs$ & $scd$ & 2 & 1 & 0 \\
                & $d$ & $s$ & 2 & 1 & $-3 e_1+ 3\sqrt{5 - 4 e ^2 _1}$ \\
                & $c$ & $c$ & 3 & 0 & $-3 e_2 + 3\sqrt{5 - 4 e ^2 _2}$ \\
                & $s$ & $d$ & 3 & 0 & $-3 e_3 + 3\sqrt{ 5 - 4 e ^2 _3}$ \\ \hline
$\ell =4$ & $\emph{1}$ & $\emph{1}$ & 0 & 4 & $e_1 h^{\emph{1}} _0 (k^2 _1) + e_3 h^{\emph{1}} _4 (k^2 _2)$\tsep{3pt} \\
                & $dc$ & $sc$ & 1 & 3 & $5e_3 - \sqrt{21 + 60e^2 _3}$ \\
                & $ds$ & $sd$ & 1 & 3 & $5e_2 - \sqrt{21 + 60e^2 _2}$ \\
                & $cs$ & $cd$ & 2 & 2 & $5 e_1 - \sqrt{21 + 60e^2 _1}$ \\
                & $\emph{1}$ & $\emph{1}$ & 2 & 2 & $e_1 h^{\emph{1}} _2 (k^2 _1) + e_3 h^{\emph{1}} _2 (k^2 _2)$ \\
                & $dc$ & $sc$ & 3 & 1 & $5e_3  + \sqrt{21 + 60e^2 _3}$ \\
                & $ds$ & $sd$ & 3 & 1 & $5e_2 + \sqrt{21 + 60e^2 _2}$ \\
                & $cs$ & $cd$ & 4 & 0 & $5e_1 + \sqrt{21 + 60e^2 _1}$ \\
                & $\emph{1}$ & $\emph{1}$ & 4 & 0 & $e_1 h^{\emph{1}} _4 (k^2 _1) + e_3 h^{\emph{1}} _0 (k^2 _2)$
\end{tabular}
\end{table}

\vspace{-3mm}

\section[Raising and lowering of quantum numbers $n_1$ and $n_2$ for spheroconal harmonic polynomials of a chosen kind and species $\ell AB n_1 n_2$]{Raising and lowering of quantum numbers $\boldsymbol{n_1}$ and $\boldsymbol{n_2}$\\ for spheroconal harmonic polynomials of a chosen kind\\ and species $\boldsymbol{\ell AB n_1 n_2}$}\label{secrln1n2}

\vspace{-1mm}

The spheroconal harmonic polynomials with their labels for angular momentum $\ell$, species $AB$ and number of nodes $n_1$ and $n_2$ can be written as the products of the pairs of matching Lam\'e polynomials
\begin{gather}
  \Psi ^{\ell AB} _{n_1 n_2} \big(\chi _1 , \chi _2 ; h^{\ell A} _{n_1}, h^{\ell B} _{n_2}\big)
  = \Lambda ^{\ell A} _{n_1} \big(\chi _1 , h^{\ell A} _{n_1}\big) \Lambda ^{\ell B} _{n_2} \big( \chi_2 , h^{\ell B} _{n_2}\big), \label{completepsi} \\
 \Lambda ^{\ell A} _{n_1} \big(\chi _1 , h^{\ell A} _{n_1}\big) = A(\chi _1)  a^A _0 \big(h^{\ell A} _{n_1} \big) \sum ^{N_{\max}} _{s=0} \bar a ^A _s \big(h^{\ell A} _{n_1}\big) \operatorname{sn}^{2s} (\chi _1) \label{completelambda}
\end{gather}
factorizing the zero power coef\/f\/icients $a ^A_0\big(h^{\ell A} _{n_1}\big)$ and using the ratio of coef\/f\/icients for the other powers
\begin{gather}
\bar a ^A _s\big(h^{\ell A} _{n_1}\big) = \frac{a ^A _s\big(h^{\ell A} _{n_1}\big)}{a ^A _0\big(h^{\ell A} _{n_1}\big)} \label{coefratio}
\end{gather}
and similarly for its companion with $A \to B$ and $n_1 \to n_2$. As established in the previous section, equations~(\ref{totalnodes}) and~(\ref{eigenvaltotal}) def\/ine the matching of quantum numbers and eigenvalues, respectively.

For chosen values of $\ell$, $A$ and $B$, the removed singularity nodes are f\/ixed in numbers~$n^A$ and~$n^B$, according to equations~(\ref{numbernodes})--(\ref{naplusnb}). Then according to equations~(\ref{nodesnp})--(\ref{nodesn2}) the number of nodes~$n^p$,~$n_1$ and~$n_2$ change in steps of~2. the inequalities in equations~(\ref{haorder}) and (\ref{hborder}) for the increasing and decreasing eigenvalues become
\begin{gather}
 h^{\ell A} _{n^A} < h^{\ell A}  _{n^A + 2} < \cdots  < h^{\ell A} _{n_1 -2} < h^{\ell A} _{n_1} < h^{\ell A} _{n_1 +2} < \cdots  < h^{\ell A} _{\ell - n^B -2} < h^{\ell A} _{\ell - n^B}   , \label{haneworder}
\end{gather}
and
\begin{gather}
 h^{\ell B} _{\ell - n^A} > h^{\ell B} _{\ell - n^A -2} > \dots  > h^{\ell B} _{ n_2 +2} > h^{\ell B} _{n_2} > h^{\ell B} _{n_2 -2} > \dots  > h^{\ell B} _{n^B +2} > h^{\ell B} _{n^B}   . \label{hbneworder}
\end{gather}

If we start from the state $\ell AB n_1 n_2$ of equation~(\ref{completepsi}) depending on the eigenvalues appearing in the middle of equations~(\ref{haneworder}) and~(\ref{hbneworder}), the state involving their neighbors to the right have raised $n_1$ by two units and lowered $n_2$ by two units with the corresponding change in the eigenvalues which remain satisfying matching conditions of equations~(\ref{totalnodes}) and~(\ref{eigenvaltotal}); the same changes operate in equations~(\ref{completelambda}) and~(\ref{coefratio}) for the eigenfunctions via the eigenvalue dependence of the expansion coef\/f\/icients. Similarly, if we had moved to the neighbors on the left: $n_1 \to n_1 -2 $ and $n_2 \to n_1 +2 $ with the complementary lowering and raising by two, with their consequences in the changes in the matching eigenvalues and the other state of the same species. Obviously, the lowering and raising actions can be extended to the right and left ends of equations~(\ref{haneworder}) and~(\ref{hbneworder}) covering all the $N^A _{\max} = N^B _{\max}$ states of the~$\ell AB$ species. Examples of these connections can be found in Tables~\ref{eigenfunc2} and~\ref{eigenenergy2}.

As a prelude to the following two sections we include a couple of purely spheroconal coordinate tools. The f\/irst one illustrates the successive species changing actions of the derivative operator on the singularity removing factors and even-power polynomials appearing in the process:

\begin{table}[h!]
\centering
\begin{tabular}{l | l | l}
   & $\frac{\partial}{\partial \chi _i}$\bsep{4pt} & $\frac{\partial ^2}{\partial \chi ^2 _i}$ \\ \hline \hline
$\emph{1}$ & 0 & 0 \\
$s$ & $cd$ & $(-k^2 _i c^2 - d^2 )s$\\
$c$ & $-sd$ &  $(-k^2 _i s^2 +d^2)c$ \\
$d$ & $-k^2 _i sc$ & $-k^2 _i (c^2 - s^2)d$ \\
$cd$ & $(-k^2 _i c^2 -d^2)s$ & $(-k^2 _i c^2 - d^2 + 2 k^2 _i s^2 )cd$ \\
$sd$ & $(-k^2 _i s^2 +c^2)c$ & $(k^2 _i s^2 - d^2 - 2 k^2 _i c^2 )sd$ \\
$sc$ & $(s^2 -c^2)d$ & $(-k^2 _i s^2 + k^2 _i c^2 + 2 k^2 _i d^2 )sc$\\
$scd$ & $-k^2 c^2 s^2 - d^2 s^2 + d^2 c^2$ & $2[k^2 _i (s^2 - c^2) + (k^2 _i s^2 - d^2) - (k^2 _i c^2 +d^2)]scd$\\
$s^2$, $c^2$, $d^2$ & $2scd \; (1,-1,- k^2 _i)$ &
\end{tabular}
\end{table}

\noindent The f\/irst derivative causes changes of species with one factor to species with the other two factors and viceversa. The species $\emph{1}$ is represented  by the top entry, and also by the bottom one; and species $dcs$ and $\emph{1}$ are also connected by their f\/irst derivatives as illustrated by the two bottom entries. The second derivative returns each species to the original one. The even-power polynomials can be reduced to polynomials in $\operatorname{sn}^2 (\chi)$ via equation~(\ref{defcndn}).

The second tool is connected with equation~(\ref{completelambda}), expressing the set of Lam\'e polynomials $\ell A n_1$ as a linear combination of the $A(\chi) \operatorname{sn}^{2s} (\chi)$ basis. Both bases have the same dimension $N^A _{\max}+1$, and the members of the latter can be expressed as linear combinations of Lam\'e polynomials. The transformation coef\/f\/icients ${\bar a} ^{A \; {\rm inv}}_{s \; n_i}$ are the elements of the inverse matrix of $\bar a^A _s (h^{\ell A} _{n_1})$.

\section[Angular momentum cartesian components connecting pairs of polynomials of the four different species with a common eigenvalue $\ell$]{Angular momentum cartesian components connecting pairs\\ of polynomials of the four dif\/ferent species with a common\\ eigenvalue $\boldsymbol{\ell}$} \label{secamtheory}

The explicit expressions for the cartesian components of the angular momentum operators in spheroconal coordinates are borrowed as equations~(A15)--(A17) in \cite{LeyKooE/MendezFragosoR:20082}:
\begin{gather}
 \hat L_x   =  \frac{-i \hbar  r^2}{{h_\chi ^2 }}\left[ {  \operatorname{dn} (\chi _1 )\operatorname{cn} (\chi _2 )\operatorname{dn} (\chi _2 )\frac{\partial }{{\partial \chi _1 }} + k_1^2 \operatorname{sn} (\chi _1 )\operatorname{cn} (\chi _1 )\operatorname{sn} (\chi _2 )\frac{\partial }{{\partial \chi _2 }}} \right]  ,\label{lx}
\\
 \hat L_y   =  \frac{-i \hbar  r^2}{{h_\chi ^2 }}\left[ - {\operatorname{cn} (\chi _1 )\operatorname{sn} (\chi _2 )\operatorname{dn} (\chi _2 )\frac{\partial }{{\partial \chi _1 }} + \operatorname{sn} (\chi _1 )\operatorname{dn} (\chi _1 )\operatorname{cn} (\chi _2 )\frac{\partial }{{\partial \chi _2 }}} \right]   ,\label{ly}
\\
 \hat L_z  =  \frac{-i \hbar  r^2}{{h_\chi ^2 }}\left[ - {k_2^2 \operatorname{sn} (\chi _1 )\operatorname{sn} (\chi _2 )\operatorname{cn} (\chi _2 )\frac{\partial }{{\partial \chi _1 }} - \operatorname{cn} (\chi _1 )\operatorname{dn} (\chi _1 )\operatorname{dn} (\chi _2 )\frac{\partial }{{\partial \chi _2 }}} \right]  . \label{lz}
\end{gather}

This section evaluates the actions of these operators on the lower $\ell AB n_1 n_2$ spheroconal harmonic polynomials of the same kind for a f\/ixed value of $\ell$, connecting with dif\/ferent species $\ell A' B' n' _1 n'_2$.

The calculations are straightforward requiring care, labor and some guidelines. The derivatives $\partial / \partial \chi _i$ act only on the f\/irst or second factor in equation~(\ref{completepsi}) for $i =1$ or $2$. The individual changes of species were already described in the f\/irst tool of the prelude, and for an initial mat\-ching species $AB$ the f\/inal ones $A'B'$ follow. Notice the appearance of the square of the angular scale factor in the three operators in equations~(\ref{lx})--(\ref{lz}). This requires that the application of the operators inside the brackets to the spheroconal harmonic polynomials provides the $A' B'$ factor and also the $h^2 _{\chi} / r^2$ factor to compensate the one in the previous line. This requirement is indeed satisf\/ied. The remaining factor can be reduced to a polynomial in even-powers of $\operatorname{sn}^2 (\chi _1)$ and $\operatorname{sn}^2 (\chi _2)$, up to $N^{A'} _{\max} = N^{B'} _{\max}$, respectively. The remaining task is to identify the harmonic linear superposition of the polynomials, making use of the linear independence and completeness of both bases in the second tool of the prelude.
Some of our results on the actions of $\hat L_x$, $\hat L_y$, $\hat L_z$ on the non-normalized spheroconal harmonic polynomials $\ell AB n_1 n_2$ for $\ell = 0,1,2,3$ are illustrated and described next.
The application of any of the operators $\hat L_i$ on the $\Lambda ^{\emph{1}} _0 \Lambda ^{\emph{1}} _0=1$ leads to the same single eigenstate with its eigenvalue zero.

\begin{table}[t]\centering
\caption{Eigenfunctions with $\ell =1$ and its resulting application of each cartesian angular momentum operator.}\label{elleq1}

\vspace{1mm}

\begin{tabular}{c | c c c}
   {} & ${\hat L_x }$ & ${\hat L_y }$\bsep{1pt} & ${\hat L_z }$  \\ \hline \hline
 $  \Lambda ^{d} _{0}  \Lambda ^{s} _{1}  $ & $ 0 $ & $-  \Lambda ^{s} _{1}  \Lambda ^{d} _{0}  $\tsep{2pt}\bsep{2pt} & $  \Lambda ^{c} _{1} \Lambda ^{c} _{0} $  \\
 $   \Lambda ^{c} _{1}  \Lambda ^{c} _{0}  $ & $  \Lambda ^{s} _{1}  \Lambda ^{d} _{0}  $\bsep{2pt} & $ 0 $ & $  - \Lambda ^{d} _{0}  \Lambda ^{s} _{1}  $  \\
$   \Lambda ^{s} _{1}  \Lambda ^{d} _{0}  $ & $  - \Lambda ^{c} _{1}  \Lambda ^{c} _{0}  $ & $  \Lambda ^{d} _{0}  \Lambda ^{s} _{1} $ & $ 0 $
\end{tabular}
\end{table}

\looseness=-1
Table~\ref{elleq1} exhibits the result of the corresponding applications on the three spheroconal har\-mo\-nics with $\ell =1$. The contents in the table should not be surprising upon recognizing the species $x$, $y$, $z$ of the successive monomials, and their equivalence with the respective cartesian harmonics.

Table~\ref{elleq2} for the $\ell =2$ polynomials contains familiar information for the unique monomial species $xy$, $xz$ and $yz$ transforming among themselves under the respective rotations, with coef\/f\/icients one in the corresponding entries. The two $\Lambda ^{\emph{1}} _0 \Lambda ^{\emph{1}} _2$ and $\Lambda ^{\emph{1}} _2 \Lambda ^{\emph{1}} _0$ binomial companions get transformed into the monomials of the other species as illustrated by the top and bottom rows in the table, with coef\/f\/icients involving the eigenvalues $h^{\emph{1}} _{n_1} (k^2 _1)$ and $h^{\emph{1}} _{n_2} (k^2 _2)$ of the matching binomials; the respective species of the monomials are complementary to that of the operators, as the reader can ascertain by moving along the top and bottom rows in the successive columns. In turn, the respective operators acting on the monomials lead to the linear superpositions of the two $\Lambda ^{\emph{1}} _{n_1} \Lambda ^{\emph{1}} _{n_2}$ binomials with more elaborate coef\/f\/icients including the common one $C^{\emph{1}}$
\begin{gather}
C^1 _{4,1} = \frac{-2 \bar a^s _1 (h^s_{3} (k^2 _2)) + \bar a^d _1 (h^d_{0} (k^2 _1) ) }{ \bar a^d _1 (h^d_{2} (k^2 _1) ) \bar a^s _1 (h^s_{3} (k^2 _2) ) - \bar a^d _1 (h^c_{0} (k^2 _1) ) \bar a^s _1 (h^s_{1} (k^2 _2) ) }   ,
\qquad C^2 _{4,1} = 1 - C^1 _{4,1} \label{c141}
\\
C^1 _{4,3} = C^1 _{4,1} \ (\textrm{by changing } d \to s)    ,
\qquad C^2 _{4,3} = -\big(1 + C^1 _{4,3}\big), \label{c141*}
\\
C^1 _{4,2} = - C^2 _{4,2} = \frac{\bar a^c _1 (h^c_{2} (k^2 _2)) + \bar a^c _1 (h^c_{1} (k^2 _1) ) }{ \bar a^c _1 (h^c_{3} (k^2 _1) ) \bar a^c _1 (h^c_{2} (k^2 _2) ) - \bar a^c _1 (h^c_{1} (k^2 _1) ) \bar a^c _1 (h^c_{0} (k^2 _2) ) }. \label{c142242}
\end{gather}

\begin{table}[t]\centering
\caption{Eigenfunctions with $\ell =2$ and its resulting application of each cartesian angular momentum operator. The common factor in the table has the value $C^{\emph{1}} = \big[ h_{0}^{\emph{1}} (k^2 _1) - h_{2}^{\emph{1}} (k^2 _1) \big] ^{-1}$.}
\label{elleq2}

\vspace{1mm}

\begin{tabular}{@{\,}c@{\,}|@{\,}c@{\,}c@{\,}c@{}}
   {} & $ {\hat L_x } $ & $ {\hat L_y } $ & $ {\hat L_z }  $\tsep{1pt} \\ \hline \hline
$    \Lambda ^{\emph{1}} _{0}  \Lambda ^{\emph{1}} _{2}  $ & $  - h_{0}^{\emph{1}} (k^2 _1)   \Lambda ^{cs} _{2}  \Lambda ^{cd} _{0} $\tsep{2pt}\bsep{2pt} & $ (h_{0}^{\emph{1}}(k^2 _1)  - h_{2}^{\emph{1}} (k^2 _2) )   \Lambda ^{ds} _{1}  \Lambda ^{sd} _{1}  $ & $ h_{2}^{\emph{1}} (k^2 _2)   \Lambda ^{dc} _{1}  \Lambda ^{sc} _{1}  $ \\
$  \Lambda ^{dc} _{1}  \Lambda ^{sc} _{1}  $ & $   \Lambda ^{ds} _{1}  \Lambda ^{sd} _{1} $ & $ -   \Lambda ^{cs} _{2}  \Lambda ^{cd} _{0} $ & $ { C^{\emph{1}} \left( \begin{array}{l}
 (2 - h_{0}^{\emph{1}} (k^2 _1) )   \Lambda ^{\emph{1}} _{2}  \Lambda ^{\emph{1}} _{0} -{}   \\
(2 - h_{2}^{\emph{1}} (k^2 _1))   \Lambda ^{\emph{1}} _{0}  \Lambda ^{\emph{1}} _{2}
 \end{array} \right)} $ \\
 $  \Lambda ^{ds} _{1}  \Lambda ^{sd} _{1} $ & $  - \Lambda ^{dc} _{1}  \Lambda ^{sc} _{1} $ & $ { 2C^{\emph{1}} \left( \begin{array}{l}
   \Lambda ^{\emph{1}} _{2}  \Lambda ^{\emph{1}} _{0}    \\
  {}- \Lambda ^{\emph{1}} _{0}  \Lambda ^{\emph{1}} _{2}
 \end{array} \right)} $ & $  \Lambda ^{cs} _{2}  \Lambda ^{cd} _{0} $ \\
$  \Lambda ^{cs} _{2}  \Lambda ^{cd} _{0} $ & $ { C^{\emph{1}} \left( \begin{array}{l}
 (2 - h_{2}^{\emph{1}} (k^2 _2) )    \Lambda ^{\emph{1}} _{2}  \Lambda ^{\emph{1}} _{0}- {}   \\
(2 - h_{0}^{\emph{1}} (k^2 _2) )   \Lambda ^{\emph{1}} _{0}  \Lambda ^{\emph{1}} _{2}
 \end{array} \right)} $ & $   \Lambda ^{dc} _{1}  \Lambda ^{sc} _{1}  $ & $ -  \Lambda ^{ds} _{1}  \Lambda ^{sd} _{1}  $ \\
$  \Lambda ^{\emph{1}} _{2}  \Lambda ^{\emph{1}} _{0}$ & $  - h_{2}^{\emph{1}} (k^2 _1)   \Lambda ^{cs} _{2}  \Lambda ^{cd} _{0} $ & $ (h_{2}^{\emph{1}} (k^2 _1)  - h_{0}^{\emph{1}} (k^2 _2) )   \Lambda ^{ds} _{1}  \Lambda ^{sd} _{1}  $\tsep{2pt} & $ h_{0}^{\emph{1}} (k^2 _2)   \Lambda ^{dc} _{1}  \Lambda ^{sc} _{1}  $
\end{tabular}

\end{table}

\begin{table}[t]\centering
\caption{Eigenfunctions with $\ell =3$ and its resulting application of each cartesian angular momentum operator. The element $C^k _{i,j}$ represents the $k$ coef\/f\/icient of the linear combination at the $i$ row and $j$ column of the table. Some examples are represented in the equations~(\ref{c141})--(\ref{c142242}).}
\label{elleq3}

\vspace{1mm}

\begin{tabular}{c | c c c}
   {} & $ {\hat L_x } $ & $ {\hat L_y } $\bsep{1pt} & $ {\hat L_z }  $ \\ \hline \hline
  $ {\Lambda _0^d \Lambda _3^s } $ & $ {2\bar a_1^d \left( {h_0^d (k_1^2 )} \right)\Lambda _2^{dcs} \Lambda _1^{scd} } $ & $ {C_{1,2}^1 \Lambda _1^s \Lambda _2^d  + C_{1,2}^2 \Lambda _3^s \Lambda _0^d } $\tsep{2pt}\bsep{2pt} & $ {C_{1,3}^1 \Lambda _1^c \Lambda _2^c  + C_{1,3}^2 \Lambda _3^c \Lambda _0^c } $  \\
  $ {\Lambda _1^c \Lambda _2^c } $ & $ {C_{2,1}^1 \Lambda _1^s \Lambda _2^d  + C_{2,1}^2 \Lambda _3^s \Lambda _0^d } $\bsep{2pt} & $ {(h_1^c (k_1^2 ) - h_2^c (k_2^2 ))\Lambda _2^{dcs} \Lambda _1^{scd} } $ & $ {C_{2,3}^1 \Lambda _0^d \Lambda _3^s  + C_{2,3}^2 \Lambda _2^d \Lambda _1^s } $ \\
 $  {\Lambda _1^s \Lambda _2^d } $ & $ {C_{3,1}^1 \Lambda _1^c \Lambda _2^c  + C_{3,1}^2 \Lambda _3^c \Lambda _0^c } $\bsep{2pt} & $ {C_{3,2}^1 \Lambda _0^d \Lambda _3^s  + C_{3,2}^2 \Lambda _2^d \Lambda _1^s } $ & $ { - 2\bar a_1^d \left( {h_2^d (k_2^2 )} \right)\Lambda _2^{dcs} \Lambda _1^{scd} }  $ \\
 $  {\Lambda _2^{dcs} \Lambda _1^{scd} } $ & $ {C_{4,1}^1 \Lambda _0^d \Lambda _3^s  + C_{4,1}^2 \Lambda _2^d \Lambda _1^s } $\bsep{2pt} & $ {C_{4,2}^1 \Lambda _1^c \Lambda _2^c  + C_{4,2}^2 \Lambda _3^c \Lambda _0^c } $ & $ {C_{4,3}^1 \Lambda _1^s \Lambda _2^d  + C_{4,3}^2 \Lambda _3^s \Lambda _0^d } $ \\
 $  {\Lambda _2^d \Lambda _1^s } $ & $ {2\bar a_1^d \left( {h_2^d (k_1^2 )} \right)\Lambda _2^{dcs} \Lambda _1^{scd} } $\bsep{2pt} & $ {C_{5,2}^1 \Lambda _1^s \Lambda _2^d  + C_{5,2}^2 \Lambda _3^s \Lambda _0^d } $ & $ {C_{5,3}^1 \Lambda _1^c \Lambda _2^c  + C_{5,3}^2 \Lambda _3^c \Lambda _0^c } $ \\
 $  {\Lambda _3^c \Lambda _0^c } $ & $ {C_{6,1}^1 \Lambda _1^s \Lambda _2^d  + C_{6,1}^2 \Lambda _3^s \Lambda _0^d } $\bsep{2pt} & $ {(h_3^c (k_1^2 ) - h_0^c (k_2^2 ))\Lambda _2^{dcs} \Lambda _1^{scd} } $ & ${C_{6,3}^1 \Lambda _0^d \Lambda _3^s  + C_{6,3}^2 \Lambda _2^d \Lambda _1^s } $ \\
$  {\Lambda _3^s \Lambda _0^d } $ & $ {C_{7,1}^1 \Lambda _1^c \Lambda _2^c  + C_{7,1}^2 \Lambda _3^c \Lambda _0^c } $\bsep{2pt} & $ {C_{7,2}^1 \Lambda _0^d \Lambda _3^s  + C_{7,2}^2 \Lambda _2^d \Lambda _1^s } $ &  ${2\bar a_1^d \left( {h_0^d (k_2^2 )} \right)\Lambda _2^{dcs} \Lambda _1^{scd} } $
\end{tabular}
\end{table}

The spheroconal harmonics with $\ell =3$ come in a single monomial of species $[dcs] [scd] = [xyz]$ and three pairs of binomials of species $x$, $y$ and~$z$. The monomial appears in the middle row of Table~\ref{elleq3}, and its companions of the other species in the rows below and above. The monomial transformed into the $x$, $y$, $z$ species in the successive columns by the respective angular momentum components; the transformed states are linear combinations of the pair of companion binomials with dif\/ferent number of nodes~$n_1$ and~$n_2$. The inverse transformations of the successive binomials leading back to the single monomial are identif\/ied in the row and column positions $(1,1)$, $(2,2)$, $(3,3)$, $(5,1)$, $(6,2)$ and $(7,3)$. The individual binomial spheroconal harmonics are transformed also into linear combinations of the pairs of binomials of species complementary to the component of the angular momentum, as an inspection of the remaining twelve positions in the table show. The coef\/f\/icients involved become more numerous, taking into account the increasing number of the states and their combinations.

\begin{figure}[h!]
\centering
\includegraphics[scale=1]{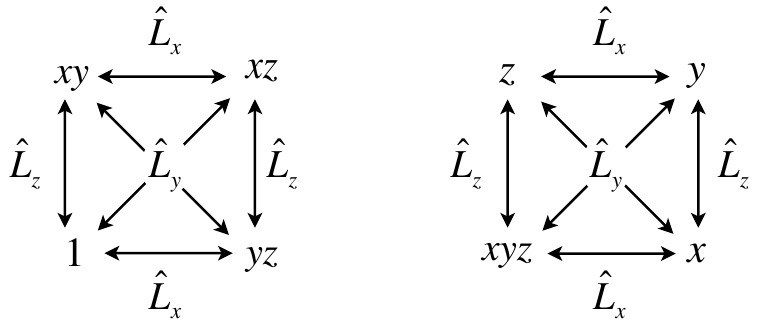}
\caption{Action of the angular momentum operators on each of the spheroconal harmonics.}
\label{diagram}
\end{figure}

We complement this section with the following diagrams in Fig.~\ref{diagram} sketching the connections among the four species, for the two kinds of even and odd $\ell$ of spheroconal harmonics, provided by the angular momentum components. The interested readers may compare them with their counterparts in~\cite{PateraJ/WinternitzP:1973}.
The coef\/f\/icients in the entries of Tables \ref{elleq1}--\ref{elleq3} are the counterpart of the raising and lowering operator coef\/f\/icients for the spherical harmonics.

\section{Linear momentum cartesian components raising and lowering\\ the angular momentum by one unit between polynomials\\ of opposite parities}\label{sectionshift}

The cartesian components of the linear momentum operator in spheroconal coordinates are
\begin{gather}
 \hat p_x    =   \operatorname{dn} (\chi _1 )\operatorname{sn} (\chi _2 ) \hat p_r \!
 -  \!\frac{i \hbar r }{{h_\chi ^2 }}\!\left[ \!{ - k_1^2 \operatorname{sn} (\chi _1 )\operatorname{cn} (\chi _1 )\operatorname{sn} (\chi _2 )\frac{\partial }{{\partial \chi _1 }}\! + \operatorname{dn} (\chi _1 )\operatorname{cn} (\chi _2 )\operatorname{dn} (\chi _2 )\frac{\partial }{{\partial \chi _2 }}} \!\right] \!,\!\!\!\!\! \label{px}
\\
 \hat p_y   =   \operatorname{cn} (\chi _1 )\operatorname{cn} (\chi _2 ) \hat p_r
 - \frac{i \hbar r}{{h_\chi ^2 }}\!\left[ { - \operatorname{sn} (\chi _1 )\operatorname{dn} (\chi _1 )\operatorname{cn} (\chi _2 )\frac{\partial }{{\partial \chi _1 }} - \operatorname{cn} (\chi _1 )\operatorname{sn} (\chi _2 )\operatorname{dn} (\chi _2 )\frac{\partial }{{\partial \chi _2 }}} \right] \! ,\!\!\! \label{py}
\\
 \hat p_z    =   \operatorname{sn} (\chi _1 )\operatorname{dn} (\chi _2 ) \hat p_r
 - \frac{i \hbar r}{{h_\chi ^2 }}\!\left[ {\operatorname{cn} (\chi _1 )\operatorname{dn} (\chi _1 )\operatorname{dn} (\chi _2 )\frac{\partial }{{\partial \chi _1 }} - k_2^2 \operatorname{sn} (\chi _1 )\operatorname{sn} (\chi _2 )\operatorname{cn} (\chi _2 )\frac{\partial }{{\partial \chi _2 }}} \right]  \!.\!\!\! \label{pz}
\end{gather}
Notice that the angular factors of $\hat p _r$  are simply the projections of the unit radial vector along the respective axes, equation~(\ref{transcoord}). The angular factor of the derivatives inside the brackets in equations~(\ref{lx})--(\ref{lz}) and (\ref{px})--(\ref{pz}) for the successive components are the same but in exchanged positions, and dif\/ferent signs in some cases. The reason for this is found in the decomposition of the $\vec p$ vector into its radial and transverse components
\begin{gather}
\hat{ \vec p} = \hat r \hat p_r  - \vec r \times \frac{{\hat{ \vec L}}}{r}  .
\end{gather}
already used in \cite{MendezFragosoR/LeyKooE:20112}. Correspondingly, also notice the presence of the factor $r / h ^2 _{\chi}$ in equations~(\ref{px})--(\ref{pz}). The actions of these on a spheroconal harmonic functions $\ell AB n_1 n_2$ lead to functions of the other kind $\ell ' A' B' n' _1 n' _2$ with $\ell ' = \ell \pm 1$, species $A' B'$ with the other parity. Consequently, the method and some of the steps and results of the previous section are useful or have their counterparts in the application of the operators of equations~(\ref{px})--(\ref{pz}) on the spheroconal harmonic polynomials $\ell AB n_1 n_2$.

At the level of each individual spheroconal coordinate, the multiplications by $d$, $c$, and $s$ of the f\/irst terms in the operators in equations~ (\ref{px})--(\ref{pz}) with the eight species of singularity removing factors $A(\chi)$ lead to the following result:

\begin{table}[h]
\centering
\begin{tabular}{l | l l l l l l l l }
& $\emph{1}$ & $s$ & $c$ & $d$ & $cd$ & $sd$ & $sc$ & $scd$ \\ \hline \hline
$d$ & $d$ & $ds$ & $dc$ & $d^2$ & $cd^2$ & $sd^2$ & $scd$ & $scd^2$ \\
$c$ & $c$ & $cs$ & $c^2$ & $cd$ & $c^2 d$ & $scd$ & $sc^2$ & $sc^2 d$ \\
$s$ & $s$ & $s^2$ & $sc$ & $sd$ & $scd$ & $s^2 d$ & $s^2 c$ & $s^2 cd$
\end{tabular}
\end{table}

\noindent with the common result that every original $A(\chi )$ is changed in kind and species. The ef\/fects are qualitatively similar and complementary to those of the f\/irst derivative in the table of the prelude at the end of Section~\ref{secrln1n2}. In fact, the union of both tables provides the sets of four species associated with the other kind of Lam\'e functions.

What has been said for $A(\chi _1)$ also holds for the companion $B(\chi _2)$, and the resulting $A' (\chi _1)$ and $B' (\chi _2)$ can be properly matched. At the level of multiplication in the f\/irst terms of equations~(\ref{px})--(\ref{pz}), their respective~$x$,~$y$ and~$z$ character has already been recognized. As a consequence of the exchange of angular factors in the derivative terms in equations~(\ref{lx})--(\ref{lz}) and \mbox{(\ref{px})--(\ref{pz})}, the kinds and species maintained in the previous section are changed in the present one. The net result is that the angular derivative terms in equations~(\ref{px})--(\ref{pz}) \mbox{acting} on the individual Lam\'e functions~$\ell A n_1$ and~$\ell B n_2$ transform them into the same kinds and species~$A' (\chi _1)$,~$B' (\chi _2)$ obtained by multiplication in the respective f\/irst terms.

We proceed to analyze the ef\/fects of the operators of equations~(\ref{px})--(\ref{pz}) on the successive Lam\'e spheroconal functions. For $\ell =0$, $\Lambda ^{\emph{1}} _0 (\chi _1) \Lambda ^{\emph{1}} _0 (\chi_2) =1$ is annihilated by the derivative terms, and transformed into
\begin{gather*}
\operatorname{dn} (\chi _1) \operatorname{sn} (\chi _2)   =   \Lambda ^d _0 (\chi _1) \Lambda ^s _1 (\chi _2) = \frac{x}{r}  , \qquad
\operatorname{cn} (\chi _1) \operatorname{cn} (\chi _2)   =   \Lambda ^c _1 (\chi _1) \Lambda ^c _0 (\chi _2) = \frac{y}{r}  ,\nonumber \\
\operatorname{sn} (\chi _1) \operatorname{dn} (\chi _2)   =   \Lambda ^s _1 (\chi _1) \Lambda ^d _0 (\chi _2) = \frac{z}{r}  ,
\end{gather*}
respectively, by the multiplication operations. The identif\/ications with the cartesian harmonics is also made. Their angular momentum of $\ell =1$ and negative parity are also recognized in contrast with the $\ell =0$ and positive parity of the original state.

Next, we illustrate the net result of the applying the derivative terms on the $\ell =1 $ states, reducing them by factoring the square of the scale factor, obtaining the following table of monomials and binomials:

\begin{table}[h!]
\centering
\caption{Actions of angular derivative terms in $\hat p_x$, $\hat p_y$, $\hat p_z$ on $\ell =1$ spheroconal harmonics.}
\label{ell1toell2}
\vspace{1mm}

\begin{tabular}{c | c c c}
   {} & ${\hat p_{x }}$ & ${\hat p_{y } }$ & ${\hat p_{z } }$  \\ \hline \hline
   $  \Lambda ^{1d} _{0}  \Lambda ^{1 s} _{1} $ & $1- \operatorname{dn}^2 (\chi _1) \operatorname{sn}^2 (\chi _2)$\tsep{2pt}\bsep{2pt} & $  \Lambda ^{2dc} _{1}  \Lambda ^{2sc} _{1} $ & $  \Lambda ^{2ds} _{1}  \Lambda ^{2sd} _{1} $  \\
 $    \Lambda ^{1c} _{1}  \Lambda ^{1c} _{0}  $ & $  \Lambda ^{2dc} _{1}  \Lambda ^{2sc} _{1}$\bsep{2pt} & $1- \operatorname{cn}^2 (\chi _1) \operatorname{cn}^2 (\chi _2) $ & $   \Lambda ^{2cs} _{2}  \Lambda ^{2cd} _{0}  $ \\
  $   \Lambda ^{1s} _{1}  \Lambda ^{1d} _{0}  $ & $   \Lambda ^{2ds} _{1}  \Lambda ^{2sd} _{1}  $ & $   \Lambda ^{2cs} _{2}  \Lambda ^{2cd} _{0} $ & $1- \operatorname{sn}^2 (\chi _1) \operatorname{dn}^2 (\chi _2)$
\end{tabular}

\end{table}

\noindent The of\/f-diagonal entries are identif\/ied as $\Lambda ^{2 dc} _1 (\chi _1) \Lambda ^{2 sc} _1 (\chi _2) = xy$, $\Lambda ^{2 ds} _1 (\chi _1) \Lambda ^{2 sd} _1 (\chi _2) = xz$ and $\Lambda ^{2 cs} _2 (\chi _1) \Lambda ^{2 cd} _0 (\chi _2) = yz$ obtained via two dif\/ferent operators acting on two dif\/ferent  initial states. They could also be obtained via the multiplication with the factors in the f\/irst terms of equations~(\ref{px})--(\ref{py}). The identif\/ication of their angular momentum $\ell =2$, and positive parity and nodal quantum numbers can be made from Table \ref{eigenenergy2}.

The diagonal entries deserve additional analysis before the lowering ef\/fect of the operators is identif\/ied, as well as the raising ef\/fects leading to the remaining two $\ell =2$ spheroconal harmonics. The three entries have been obtained via three dif\/ferent operators acting on three dif\/ferent initial functions. Of course, the ``ones'' correspond to the monopolar harmonic to which we go back by the lowering ef\/fect on the dipole harmonics. On the other hand, the other terms are identif\/ied as the squares of the projections of the radial unit vector along the coordinate axes. Their sum is ``one'' coinciding again with the lowered $\ell = 0$ state. Of the three terms, only two are linearly independent, since their sum is f\/ixed. This suggests that the three terms can be combined in two linearly independent combinations to be identif\/ied with the missing $\Lambda ^{2 \emph{1}} _0 (\chi _1) \Lambda ^{2 \emph{1}} _2 (\chi _2)$ and $\Lambda ^{2 \emph{1}} _2 (\chi _1) \Lambda ^{2 \emph{1}} _0 (\chi _2)$, $\ell =2 $ spheroconal harmonics.

The analysis can be made more transparent by recognizing that $x^2$, $y^2$ and $z^2$ are equivalent to them upon multiplication by $r^2$. Let us consider the linear combination
\begin{gather}
\Phi (x,y,z) = \alpha x^2 + \beta y^2 + \gamma z^2 . \label{generalphi}
\end{gather}
It is advantageous to choose it as harmonic
\begin{gather}
\nabla ^2 \Phi = 2\alpha + 2\beta + 2\gamma = 0 , \label{nablaapplied}
\end{gather}
because in such a case we are sure that their connection with the spheroconal harmonics is linear. In fact, from equation~(\ref{nablaapplied}) we get
\begin{gather*}
\beta = - \alpha - \gamma  ,
\end{gather*}
and rewrite equation~(\ref{generalphi}) in the separable form in spheroconal coordinates
\begin{gather}
\Phi (r, \chi _1 , \chi _2) = r^2 \left[ \alpha \operatorname{dn}^2 (\chi _1) \operatorname{cn}^2 (\chi _2) - (\alpha + \gamma) \operatorname{cn}^2 (\chi _1) \operatorname{cn}^2 (\chi _2) + \gamma \operatorname{sn}^2 (\chi _1) \operatorname{dn}^2 (\chi _2) \right]. \label{nablappsphero}
\end{gather}
The question to be answered is which values of the coef\/f\/icients $\alpha$ and $\gamma$ combining the diagonal angular entries lead to the $\ell =2$ spheroconal harmonics of species $\emph{1}\;\emph{1}$.

\looseness=-1
The answers are obtained by rewriting equation~(\ref{nablappsphero}) as polynomials in even-powers of $\operatorname{sn} (\chi _1)$ and $\operatorname{sn} (\chi _2)$ and comparing them with the target polynomials from Table \ref{eigenfunc2}. Their explicit forms~are
\begin{gather*}
\alpha _{0\;2} = - \frac{1}{2} + \frac{1}{4} h^{\emph{1}} _0 \big(k^2 _1\big) -  \frac{1}{4} h^{\emph{1}} _2 \big(k^2 _2\big)  , \qquad
\gamma _{0\;2} = - \frac{1}{2} - \frac{1}{4} h^{\emph{1}} _0 \big(k^2 _1\big) +  \frac{1}{4} h^{\emph{1}} _2 \big(k^2 _2\big)  ,
\end{gather*}
and
\begin{gather*}
\alpha _{2\;0} = - \frac{1}{2} + \frac{1}{4} h^{\emph{1}} _2 \big(k^2 _1\big) -  \frac{1}{4} h^{\emph{1}} _0 \big(k^2 _2\big)  , \qquad
\gamma _{2\;0} = - \frac{1}{2} - \frac{1}{4} h^{\emph{1}} _2 \big(k^2 _1\big) +  \frac{1}{4} h^{\emph{1}} _0 \big(k^2 _2\big) ,
\end{gather*}
def\/ining the proportions of the presence of the respective spheroconal harmonics in the diagonal entries in the Table~\ref{ell1toell2}. We remind the reader that the analysis has been made for non-normalized polynomials. In any case, the coef\/f\/icients above are the counterpart of the familiar spherical Clebsch--Gordan coef\/f\/icients. Work on the raising and lowering from $\ell = 2 ,3, \dots $ is in process.

\section{Discussion\label{secdiscu}}

Some of our works on the rotations of asymmetric molecules and the Lam\'e spheroconal harmonics have taken advantage of their connections with their spherical and cartesian counterparts~\mbox{\cite{LeyKooE/MendezFragosoR:20081,MendezFragosoR/LeyKooE:20112}}, and others have been developed strictly within the spheroconal formalism \cite{LeyKooE/MendezFragosoR:20082,MendezFragosoR/LeyKooE:2010,MendezFragosoR/LeyKooE:2011,MendezFragosoR/LeyKooE:20112}. The analysis and identif\/ication of the three sets of ladder operators in this article has been purposefully implemented in the same formalism, avoiding the spherical and cartesian crutches.

The connections among Lam\'e spheroconal polynomials of a given kind and species, and dif\/ferent numbers of nodes have been explicitly recognized in Section~\ref{secrln1n2}. They share the same form and degree, dif\/fering in their number of nodes $n_i$; in contrast with the classical polynomials having a number of nodes equal to their degree. For the spheroconal harmonics constructed by the product of Lam\'e polynomials with matching number of nodes and eigenvalues, equations~(\ref{totalnodes}) and~(\ref{eigenvaltotal}), their complete set is connected by their complementary raising of $n_1$ and lowering of $n_2$. The inversion of equation~(\ref{completelambda}) provides the basis of the powers of $\operatorname{sn}^2 (\chi)$ as a linear combination of the set of corresponding Lam\'e polynomials, as the counterpart of the known relationships in the classical polynomials.

Our other contribution in this volume of SIGMA \cite{LeyKooE/GuoHuaS:2012} contains another example of polynomial eigenfunctions, for the Hydrogen atom conf\/ined by a dihedral angle in prolate spheroidal coordinates, evaluated by diagonalization of tridiagonal f\/inite size matrices, and their connections by complementary changes in the numbers of prolate spheroidal and hyperboloidal nodes.

\looseness=-1
The shifting actions of the $\hat L_x$, $\hat L_y$, $\hat L_z$ operators among the spheroconal harmonics of the four species for a chosen value $\ell$, as illustrated in Tables \ref{elleq2} and \ref{elleq3}, allow the generation of the corresponding coef\/f\/icients in the purely spheroconal formalism. They are the counterpart of the familiar raising and lowering coef\/f\/icients $C_{\pm} (\ell , m )$ for the spherical harmonics, and even closer for the spherical harmonics with def\/inite parities. The relationships so far are for the non-normalized coef\/f\/icients, but work is under way to identify and incorporate the normalization coef\/f\/icients.

The $\ell$ raising and lowering actions of $\hat p_x$, $\hat p_y$ and $\hat p_z$ among spheroconal harmonics of neigh\-bo\-ring kinds and species has been illustrated for the lower states, but the systematics had already been recognized in \cite{MendezFragosoR/LeyKooE:20112}. While writing the latter, several works on related topics \cite{LiuQH/XunDM/ShanL:2010,SunGH/DongSH:2010,SunGH/DongSH:2011} allowed us to recognize the complete radial and angular raising and lowering actions for the free particle in spherical coordinates. Now we can also extend it to spheroconal coordinates.

Pi\~na identif\/ied four step $\Delta \ell = 4$ ladder operators for spheroconal harmonics of the same species for the most asymmetric molecules \cite{PinaE:1994}. We are investigating their possible connection with the application of the $\hat p _i$ operators four times in succession leading to the same states. In such a case the results are valid for any asymmetry distribution.

The familiarity with cartesian, spherical and spheroconal harmonics is decreasing when written in this order. The works \cite{LeyKooE/MendezFragosoR:20081,PateraJ/WinternitzP:1973} used the spherical basis as the tool to evaluate the spheroconal harmonics. In the process the interbasis expansion coef\/f\/icients are generated for the original problem, and automatically also for the inverse problem. The specif\/ic example at the end of Section \ref{sectionshift} shows the spheroconal quadrupole composition of the cartesian one in equation~(\ref{generalphi}).

As a complement to the comment in the third paragraph of this section, we include the following related references with their respective comments. The contents in \cite{AquilantiV/CaligianaA/CavalliS:2003} for the free Hydrogen atom in conf\/iguration space overlap with ours in \cite{LeyKooE/GuoHuaS:2012}, with emphasis on the rarely treated spheroidal sets, and in the present contribution, on the properties of spheroelliptic orbitals, which have been so far practically ignored. The related work \cite{AquilantiV/CaligianaA/CavalliS/ColettiC:2003} also deals with the free Hydrogen atom in momentum space focusing on the hyperspherical harmonic elliptic Sturmian basis set in $S^3$. The earlier work \cite{GroscheC/KarayanKhG/PogosyanGS/SissakianAN:1997} deals with the free quantum motion on the three-dimensional sphere $S^3$ ellipso-cyllindrical coordinates, distinguishing between prolate and oblate elliptic coordinates. On the other hand the more recent work \cite{AquilantiV/TonzaniS:2004} deals with the use of the hyperspherical elliptic coordinates and harmonic basis sets in the f\/ifth dimensional sphere $S^5$ for the study of the three-body problem in quantum mechanics. Its classical astronomical counterpart \cite{PinaE/JimenezLaraL:2002} uses the relative radial coordinate of the f\/irst two particles and the radial coordinate of the third particle relative to the center of mass of the f\/irst two, instead of the hyper-radius  and one of the f\/ive angular coordinates in $S^5$. It is also pertinent to mention Sasaki's contribution in the Symposium, ``Exactly solvable quantum mechanics and inf\/inite families of multi-indexed orthogonal polynomials'', and to cite \cite{OdakeS/SasakiR:2011, OdakeS/SasakiR:2008}, involving Hamiltonians with tridiagonal matrix representations. The raising and lowering operators in our Section~\ref{secrln1n2} have their counterparts for the respective eigenfunctions in the references of this paragraph.

\subsection*{Acknowledgements}

The authors express their appreciation to the organizers of the Symposium and editors of this Volume of SIGMA on ``Superintegrability, Exact Solvability, and Special Functions" for their invitations to participate in both. The authors acknowledge the f\/inancial support for this work by Consejo Nacional de Ciencia y Tecnolog\'ia, SNI-1796.

\pdfbookmark[1]{References}{ref}

\LastPageEnding

\end{document}